\documentclass[12pt,amsmath,aps,amssymb,showpacs]{revtex4}
\usepackage{graphicx}
\usepackage{dcolumn}
\usepackage{bm}

\begin{document}

\title{Classical and quantum Cosmology of the S\'aez-Ballester theory}
\author{J. Socorro}
\email{socorro@fisica.ugto.mx}
\author{M. Sabido}
\email{msabido@fisica.ugto.mx}
\author{L. Arturo Ure\~na-L\'opez}
\email{lurena@fisica.ugto.mx}
\affiliation{Departamento de F\'isica, DCI, Campus Le\'on, Universidad
  de Guanajuato, C.P. 37150,  Guanajuato, M\'exico}

\begin{abstract}
  We study the  generalization of the S\'aez-Ballester theory applied
  to a flat FRW cosmological model. Classical exact solutions up to
  quadratures are easily obtained using the Hamilton-Jacobi
  approach.  Contrary to claims in the
  specialized literature, it is shown that the S\'aez-Ballester theory
  cannot provide a realistic solution to the dark matter problem of
  Cosmology.  Furthermore the quantization procedure of the theory  can be simplified by 
  reinterpreting the theory in the Einstein frame, where the scalar field can be interpreted as part of the matter content of the theory, in this approach, exact solutions are also found for the Wheeler-DeWitt
  equation in the quantum regime.
\end{abstract}

\maketitle

\section{Introduction}
The inclusion of scalar fields into homogeneous cosmologies is a
typical practice to study different scenarios, such as inflation, dark
matter, and dark energy\cite{Copeland:2006wr}. However, since the
early seventies, the problem exists of finding the appropriate sources
of matter and its corresponding Lagrangian to solve an  specific
scenario\cite{ryan1,ryan}.

In this respect, Saez and Ballester (SB)\cite{s-b} formulated a
scalar-tensor theory of gravitation in which the metric is coupled to
a dimensionless scalar field in order to solve the so-called missing
matter problem in Cosmology. Some works about the classical regime are
already present in the
literature\cite{singh,shri,mohanty,singh-shri}. In particular, in ref.
\cite{singh-shri} the authors consider the coupling parameter time-dependent
and take a particular ansatz for mathematical convenience for solving the field equations.

In spite of a the dimensionless character of the scalar field, an
antigravity regime appears, and this fact has been used to suggest a
new possible way to solve the missing matter problem in non-flat FRW
cosmologies. On the other hand, the quantization program of the theory
has yet to be made.

In this paper, we shall study a generalization of the SB theory and
transform it into a conventional tensor theory, where the
dimensionless scalar field is interpreted as an exotic matter. We found the general 
behaviour for the kinetic scalar field
dependent to the scale factor of the universe, but the behaviour corresponds to stiff 
matter and  not for a dust universe, then the missing matter problem is not solved. 

 With respect to the quantization program, in this approach we can construct the 
quantization program of the theory using
the usual ADM formalism\cite{ryan1}.  Also, we can  in principle quantize the theory 
following the Loop Quantum Cosmoloy program.

In this work, we shall use this formulation  to obtain classical and quantum  solutions 
in quadratures, for the flat barotropic FRW cosmology,  including a cosmological
term  $\lambda$. 

 The paper is arranged as follows, In section II we write the generalization S\'aez-Ballester
 formalism in the usual manner, 
that is, we calculate the
corresponding energy-momentum tensor to the scalar field and give the equivalent lagrangian density.
 Next, we proceed to obtain
the corresponding canonical lagrangian ${\cal L}_{can}$ to a flat FRW universe through the lagrange transformation, 
we calculate the classical hamiltonian, we also present solutions to some models. In section III, 
using the transformation and the Hamiltonian constraint 
${\cal H}$, f we find the Wheeler-DeWitt (WDW) equation of the corresponding cosmological model 
under study. Section IV  is devoted to conclusions and outlook.

\section{Generalized Saez-Ballester theory}
The simplest generalization of the S\'aez-Ballester theory\cite{s-b}
with a cosmological term is  
\begin{equation}
  {\cal L}_{geo} = \left( R- 2 \lambda - F(\phi) \phi_{,\gamma}
    \phi^{,\gamma}\right) \, , \label{lagrangian}
\end{equation}
where $R$ the scalar curvature, $\phi^{,\gamma} = g^{\gamma \alpha}
\phi_{,\alpha}$, and $F(\phi)$ is a dimensionless and arbitrary
functional of the scalar field. According to common wisdom, the
Lagrangian~(\ref{lagrangian}) would correspond to a scalar field
theory without scalar potential but with an exotic kinetic term. 

The complete action is then
\begin{equation}
  I = \int_{\Sigma} \sqrt{-g}({\cal L}_{geo} + {\cal L}_{mat}) \, d^4x
  \, , \label{action}
\end{equation}
where we have included a matter Lagrangian ${\cal L}_{mat}$, and $g$
is the determinant of metric tensor. The field equations derived from
the above action are
\begin{subequations}
\begin{eqnarray}
  G_{\alpha \beta} + g_{\alpha\beta} \lambda - F(\phi) \left(
    \phi_{,\alpha} \phi_{,\beta} - \frac{1}{2} g_{\alpha \beta}
    \phi_{,\gamma} \phi^{,\gamma} \right) &=& 8\pi G T_{\alpha \beta}
  \, , \label{efe} \\
  2F(\phi) \phi^{,\alpha}_{\,\,;\alpha} + \frac{dF}{d\phi}
  \phi_{,\gamma} \phi^{,\gamma} &=& 0 \, , \label{fe}
\end{eqnarray}
\end{subequations}
in which $G$ is the gravitational constant, and a semicolon means
covariant derivative.

The same set of equations(\ref{efe},\ref{fe}) is obtained if we consider the
scalar field $\phi$ as part of the matter budget, i.e. say 
$\rm {\cal L}_{\phi}=\rm F(\phi) g^{\alpha \beta}\phi_{,\alpha} \phi_{,\beta}$. In this new line of reasoning,  action (\ref{action}) can be rewritten as a geometrical part (Hilbert-Einstein with $\Lambda$) and matter content (usual matter plus a term that corresponds to the scalar field component of S\'aez-Ballester theory),
\begin{equation}
I = \int_{\Sigma} \sqrt{-g} \left( R- 2\lambda + {\cal L}_{mat} +
  {\cal L}_\phi \right) \, d^4x \, . \label{action1}
\end{equation}

Even though the philosophy is different to that of the original SB
theory, the similarity of the latter to a standard scalar field theory
at the classical level will help us to infer the correspondence
quantum formulation.  We expect the quantum picture will also be the
correct one for the SB theory, as all the formulation is based upon
the same (classical) Hamiltonian constraint.

Using this action we obtain the classical Hamiltonian of the
generalized SB theory for a Friedmann-Robertson-Walker background. Let
us start with the line element for a homogeneous and isotropic
universe,
\begin{equation}
  ds^2= -N^2(t) dt^2 + a^2(t) \left[ \frac{dr^2}{1-\kappa r^2} + r^2
    d\Omega^2 \right] \, , \label{frw}
\end{equation}
where $a(t)$ is the scale factor, $N(t)$ is the lapse function, and
$\kappa$ is the curvature constant that can to take the values $0$,
$1$ and $-1$, for flat, closed and open universe, respectively. The
total Lagrangian density then reads
\begin{equation}
  {\cal L} = \frac{6\dot a^2 a}{N} - 6\kappa N a + \frac{F(\phi)
    a^3}{N} \dot \phi^2 + 16\pi G N a^3 \rho - 2N a^3 \lambda \,
  , \label{frw-lagrangian}
\end{equation}
where $\rho$ is the matter energy density; we will assume that it
complies with a barotropic equation of state of the form $p=\gamma
\rho$, where $\gamma$ is a constant. The conjugate momenta are
obtained from
\begin{eqnarray}
  \Pi_a &=& \frac{\partial {\cal L}}{\partial \dot a} = \frac{12a\dot
    a}{N}, \qquad \rightarrow \qquad \dot a = \frac{N\Pi_a}{12 a} \, ,
  \nonumber \\
  \Pi_\phi &=& \frac{\partial {\cal L}}{\partial \dot \phi} = \frac{2F
    a^3 \dot{\phi}}{N} \, , \qquad \rightarrow \qquad \dot{\phi} =
  \frac{N\Pi_\phi}{2F a^3 } \, . \label{momentas}
\end{eqnarray}
From the  canonical form of  the Lagrangian density~(\ref{frw-lagrangian})  and the 
 solution for the barotropic fluid equation of motion we find  the Hamiltonian
density for this theory
\begin{equation}
  {\cal H} = \frac{a^{-3}}{24} \left[ a^2 \Pi_a^2 + \frac{6}{F(\phi)}
    \Pi_\phi^2 + 144 \kappa a^4 + 48a^6 \lambda -384 \pi G \rho_\gamma
    a^{3(1-\gamma)} \right] , \label{hamiltonian}
\end{equation}
where $\rho_\gamma$ is an integration constant.


\subsection{Classical solutions for flat FRW}
Using the transformation $\Pi_q=\frac{d S_q}{d q}$, the
Einstein-Hamilton-Jacobi corresponding to Eq.~(\ref{hamiltonian}) is
\begin{equation}
  a^2 \left( \frac{d S_a}{d a} \right)^2 + \frac{6}{F(\phi)}
  \left(\frac{d S_\phi}{d \phi} \right)^2 + 48a^6 \lambda - 384 \pi G
  \rho_\gamma a^{3(1-\gamma)} = 0 \, ,.
\end{equation}
The EHJ equation can be further separated in the equations
\begin{eqnarray}
  \frac{6}{F(\phi)} \left( \frac{d S_\phi}{d \phi} \right)^2 &=& \mu^2
  \, , \label{pphi} \\
  a^2 \left( \frac{d S_a}{d a} \right)^2  + 48a^6 \lambda - 384 \pi G
  \rho_\gamma a^{3(1-\gamma)} &=& -\mu^2 \, ,  \label{aa}
\end{eqnarray}
where $\mu$ is a separation constant. With the help of
Eqs.~(\ref{momentas}), we can obtain the solution up to quadratures of
Eqs.~(\ref{pphi}) and~(\ref{aa}),
\begin{subequations}
\begin{eqnarray}
  \int \sqrt{F(\phi)} \, d\phi &=& \frac{\mu}{2\sqrt{6}}\int a^{-3}
  (\tau) \, d\tau \, , \label{phi-new} \\
  \Delta \tau &=& \int \frac{a^2da}{\sqrt{\frac{8}{3}\pi G \rho_\gamma
      a^{3(1-\gamma)} - \frac{\lambda}{3}a^6 - \nu^2}} \,
  , \label{aa-new}
\end{eqnarray}
\end{subequations}
with $\nu=\frac{\mu}{12}$

Eq.~(\ref{phi-new}) readily indicates that
\begin{equation}
  F(\phi) \dot \phi^2 = 6\nu^2 a^{-6}(\tau) \, , \label{missing}
\end{equation}
despite of the particular form of the functional $F(\phi)$. Also, this structure is directly obtained for this model
solving the equation (\ref{fe}). Moreover,
the matter contribution of the SB scalar field to the rhs of the
Einstein equations would be
\begin{equation}
  \rho_\phi = \frac{1}{2} F(\phi) \dot{\phi}^2 \propto a^{-6} \, .
\end{equation}
That is, the contribution of the scalar field is the same as that of
stiff matter with a barotropic equation of state $\gamma = 1$.

This is an interesting result, since the original SB theory was
thought of as a form to solve the missing matter problem of Cosmology,
now generically called the dark matter problem; to solve the latter,
one needs a fluid behaving as dust with $\gamma = 0$. It is surprising that
such a general result remain unnoticed until now in the literature
about SB.  

Also, that we have identified the general evolution of the scalar
field with that of a stiff fluid means that the Eq.~(\ref{aa-new})
can be integrated separately without a complete solution for the
scalar field. For completeness, we give below a compilation of exact
solutions in the case of the original SB theory.

If $F(\phi)=\omega \phi^m$, then we have two cases that correspond to
$m=-2$ and $m \neq -2$; the general solution for the scalar field is
\begin{equation}
  \phi=\left\{
    \begin{tabular}{lr}
      $Exp \left[ \frac{6\nu}{\sqrt{6\omega}} \int a^{-3} (\tau) d\tau
      \right]$ & \qquad m = -2 \\
      $\left[ \frac{2\nu(m+2)}{\sqrt{6\omega}} \int a^{-3} (\tau)
        d\tau \right]^{\frac{2}{m+2}}$ & \qquad $m \not=-2$ \\
    \end{tabular}
  \right. \label{sol-phi}
\end{equation}
which can be completely integrated once the time dependence of the
scale factor $a$ has been resolved.

\begin{itemize}
\item Stiff plus a cosmological constant, $\gamma=-1$. The master
  equation become
  \begin{equation}
    \Delta \tau = \int \frac{a^2da}{\sqrt{b_{-1} a^6 - \nu^2}} \, ,
  \end{equation}
where $b_{-1} = \frac{8}{3}\pi G \rho_{-1} - \frac{\lambda}{3}$, whose
solution is
\begin{equation}
  \Delta \tau = \frac{1}{3\sqrt{b_{-1}}} \, Ln \left[b_{-1} a^3 +
    \sqrt{b_{-1}} \sqrt{b_{-1} a^6 - \nu^2} \right] \, .
\end{equation}
The volume function is then
\begin{equation}
  a^3 = \frac{1}{2b_{-1}} \left(e^{3\sqrt{b_{-1}} \, \Delta \tau} +
    b_{-1} \nu^2 e^{-3\sqrt{b_{-1}} \, \Delta \tau} \right) \, ,
\end{equation}
whereas that of the scalar field is
\begin{equation}
  \phi = \left\{
    \begin{tabular}{lr}
      $Exp \left[ \frac{4}{\sqrt{6\omega}} \arctan \left( \frac{Exp [ 3
            \sqrt{b_{-1}} \Delta \tau]}{\nu \sqrt{b_{-1}}} \right)
      \right]$ & \qquad $m = -2$ \, ; \\
      $\left[ \frac{2(m+2)}{\sqrt{6\omega}} \arctan\left( \frac{Exp [3
            \sqrt{b_{-1}} \Delta \tau]}{\nu \sqrt{b_{-1}}} \right)
      \right]^{\frac{2}{m+2}}$ & \qquad $m \not= -2$ \, . \\
\end{tabular}
\right. \label{sol-phi-tau}
\end{equation}

For the case  $\gamma=1$ the same solutions are found and only a redefinition of the constants is needed.
\item Stiff plus a cosmological constant plus dust, $\gamma=0$. In
  this case the master equation becomes
\begin{equation}
  \Delta \tau =\int \frac{a^2da}{\sqrt{\frac{8}{3}\pi G \rho_0 a^3 -
      \frac{\lambda}{3} a^6 - \nu^2}}
\end{equation}
whose solution is
\begin{equation}
  \Delta \tau = \frac{1}{\sqrt{3|\lambda|}} \, Ln \left[ \frac{b_0 +
      \frac{2 |\lambda|}{3} a^3}{\sqrt{\frac{|\lambda|}{3}}} + 2
    \sqrt{b_0 a^3 + \frac{|\lambda|}{3} a^6 - \nu^2} \right] \, ,
\end{equation}
with $|\lambda|>0$ and $\rm b_0=\frac{8}{3}\pi G \rho_0$. The volume
function is now
\begin{equation}
  a^3 = \frac{3}{4\sqrt{3|\lambda|}} e^{-\sqrt{3|\lambda| }\tau}
  \left[ 4\nu^2 + \left( e^{\sqrt{3|\lambda| }\tau} -
      \frac{3b_0}{\sqrt{3|\lambda|}} \right)^2 \right] \, .
\end{equation}
In this way, the solution for the field $\phi$ is
\begin{equation}
  \phi=\left\{
    \begin{tabular}{lr}
      $Exp \left[ \frac{4}{\sqrt{6\,\omega}} \arctan \left(
          \frac{\sqrt{3|\lambda|} Exp[\sqrt{3|\lambda|}\Delta \tau] -3
            b_0}{2\nu\sqrt{3|\lambda|}} \right) \right]$ & \qquad $m =
      -2$ \, ;\\
      $\left[ \frac{4(m+2)}{\sqrt{6\omega}} \arctan \left(
          \frac{\sqrt{3|\lambda|} Exp[\sqrt{3|\lambda|} \Delta \tau] -
            3b_0}{2 \nu \sqrt{3|\lambda|}} \right)
      \right]^{\frac{2}{m+2}}$ & \qquad $m \not= -2$ \, . \\
    \end{tabular}
\right. \label{sol-phi-tau-1}
\end{equation}
\end{itemize}

The classical solution when $\rm F(\phi)=we^{m\phi}$ have the following
structure
\begin{equation}
\rm \phi(\tau)=\frac{2}{m} Ln\left[\frac{m}{2}\sqrt{\frac{6\nu^2}{w}} \int a^{-3}(\tau) d\tau + e^{\frac{m}{2}\phi_0}\right],
\end{equation}
where the integration value must be consider the last calculations over the scale factor.

The solutions above were checked to comply with the Einstein field
equations encoded in equations (\ref{fe}), using the REDUCE 3.8
package.

\section{Quantum FRW cosmological model}
One of the open problem of SB is the lack of a quantum model, in this section using the generalization 
of the ideas presented in the previos sections we use canonical quantuization.
By the usual representation for the momenta operators $\rm \Pi_q=-i\frac{\partial}{\partial q} $, $(\hbar=1),$
including the factor ordering problem in the $a$ and $\phi$ variables,
we obtain the Wheeler-DeWitt equation
\begin{equation}
\rm \left[-a^2\frac{\partial^2}{\partial a^2}-qa\frac{\partial}{\partial a}-\frac{6}{F(\phi)} 
\frac{\partial^2}{\partial \phi^2}
-\frac{6s}{F(\phi)}\phi^{-1}\frac{\partial}{\partial \phi}+144\kappa a^4+48a^6 \lambda - 
384\pi G \rho_\gamma a^{3(1-\gamma)} \right] 
\Psi=0, \label{wdw}
\end{equation}
where q and s are real constants that measures the ambiguity in the factor ordering in the 
operators $\Pi_a$ and $\Pi_\phi$, 
$\Psi$ is the wave function for this cosmological model. Employing the variables separation 
 method, $\Psi(a,\phi)={\cal A}(a){\cal B}(\phi)$, 
(\ref{wdw}) gives the set of equations
\begin{eqnarray}
\rm -a^2\frac{d^2{\cal A}}{da^2}-qa\frac{\partial {\cal A}}{\partial a}+
\left(144\kappa a^4+48a^6 \lambda -384 \pi G \rho_\gamma a^{3(1-\gamma)}-\mu^2\right) {\cal A}&=&0,\label{a}\\
\rm \phi \frac{d^2 {\cal B}}{d\phi^2}+s \frac{ d {\cal B}}{d \phi}-\frac{\mu^2}{6}\phi F(\phi) {\cal B}&=&0.\label{phi}
\end{eqnarray}

The equation (\ref{a}) does have not a general solution for any $\kappa$, then we solve for flat
 case and the particular 
values in the $ \gamma$ parameter.
When $\gamma=-1$, the exact solution is
\begin{equation}
\rm {\cal A}(a)=a^{\frac{1-q}{2}}\, Z_\nu\left( \frac{\sqrt{b}}{3}a^3\right), \label{a-solution}
\end{equation}
where $\nu=\frac{1}{6}\sqrt{(1-q)^2-4\mu^2}$ and $b=384\pi G \rho_{-1}-48\lambda$. We can see that when $b>0$, 
the generic Bessel function $Z_\nu\to J_\nu$,
and when $b<0$, $Z_\nu\to (K_\nu, I_\nu)$

Other soluble case is when $\gamma=1$, the solution is the same, and the changes appear in 
the constants $\mu^2\to 384\pi G\rho_1+\mu^2$ and $b=-48\lambda$. In this form, we obtain the exact
 solution to the wave function $\Psi(a,\phi)$
in this theory.

For solve the equation (\ref{phi}), we apply this approach at S\'aez-Ballester theory. The case when $m\not=-2$ 
\cite{polyanin} is written in term of generic Bessel function $\rm Z_\eta$ as
\begin{equation}
\rm B(\phi)= 
 c \phi^{\frac{1-s}{2}} Z_\eta \left( \frac{2\sqrt{-\xi}}{m+2} \phi^{\frac{m+2}{2}}\right),
\end{equation} 
where c is a integration constants, and $\eta=\frac{1-s}{m+2}$, $\xi=\frac{\mu^2\omega}{6}$.
 Also, we can see that the
 generic Bessel function $\rm Z_\eta \to J_\eta$ when $\omega<0$, or $\rm (K_\eta, I_\eta)$ when $\omega>0$. 

 We can build the wave packet, introducing the continuum parameters $\eta$ and $\nu$ as
\begin{equation}
\rm \Psi_{\eta\nu}=\int_\eta \int_\nu {\cal F}(\eta){\cal G}(\nu) 
\phi^{\frac{1-s}{2}} Z_\eta \left( \frac{2\sqrt{-\xi}}{n+2} \phi^{\frac{n+2}{2}}\right)
a^{\frac{1-q}{2}}\, Z_\nu\left( \frac{\sqrt{b}}{3}a^3\right) d\eta d\nu
\end{equation}

For particular values in the constant $m$, the exact solutions are very simple. For instant when $m=-2$, 
we have the Euler equation who
solution is
\begin{equation}
\rm B(\phi)= \phi^{\frac{1-s}{2}}\left\{
\begin{tabular}{lr}
$\rm  \left[c_1 \phi^\alpha + c_2 \phi^{-\alpha} \right]$ & \qquad $(1-s)^2> 4b$ \\
$\rm  \left[c_1  + c_2 Ln \phi \right]$ & \qquad $(1-s)^2= 4b$ \\
$\rm  \left[c_1 sin(\alpha Ln \phi) + c_2 cos(\alpha Ln(\phi)) \right]$ & \qquad $(1-s)^2< 4b$
\end{tabular}
\right. 
\end{equation} 
with $\alpha=\frac{1}{2}\sqrt{(1-s)^2-4b}$ and $b=-\frac{\omega\mu^2}{6}$.

When $m=-6$ and $s=-1$, making the transformations $z=\phi^{-2}$ and $B=\frac{u}{z}$, leads to a 
constant coefficient
linear equation, (\ref{phi}) is transformed
to $4\frac{d^2u}{dz^2}-\frac{\mu^2 \omega}{6}u=0$ who exact solutions becomes
\begin{equation}
u(z)=\left\{
\begin{tabular}{lr}
$\rm c_1 \, sinh\left(\sqrt{\frac{\mu^2\omega}{24}}z\right) + c_2 \,cosh\left(\sqrt{\frac{\mu^2\omega}{24}}z\right)$&\qquad $\omega>0$\\
$\rm c_1 \, sin\left(\sqrt{\frac{\mu^2\omega}{24}}z\right) + c_2 \,cos\left(\sqrt{\frac{\mu^2\omega}{24}}z\right)$&\qquad $\omega<0$\\
\end{tabular} \right.
\end{equation}
in the original variables
\begin{equation}
{\cal B}(\phi)= \phi^2\left\{
\begin{tabular}{lr}
$\rm c_1 \, sinh\left(\sqrt{\frac{\mu^2\omega}{24}}\frac{1}{\phi^2}\right) + c_2 \,cosh\left(\sqrt{\frac{\mu^2\omega}{24}}\frac{1}{\phi^2}\right)$&\qquad $\omega>0$\\
$\rm c_1 \, sin\left(\sqrt{\frac{\mu^2\omega}{24}}\frac{1}{\phi^2}\right) + c_2 \,cos\left(\sqrt{\frac{\mu^2\omega}{24}}\frac{1}{\phi^2}\right)$&\qquad $\omega<0$\\
\end{tabular} \right.
\end{equation}

\section{conclusions}
We studied the generalization of the S\'aez-Ballester theory by
including a dimensionless functional of the scalar field
$F(\phi)$. The classical dynamics of the theory were obtained from the
corresponding classical Lagragian and Hamiltonian densities; the
solutions were in turn given up to quadratures.

One general result here is that the evolution of the scale factor of
the Universe does not depend upon the particular form of the
functional $F(\phi)$; actually, the contribution of the scalar field
in the SB theory is that of perfect fluid with a stiff (barotropic)
equation of state. If any, its contribution to the matter budget of
the Universe is only relevant at early times.

A separate conclusion is that the SB, whether in its original form as
given in Ref.\cite{s-b} or in the generalized case studied here,
cannot be an answer to the dark matter riddle of Cosmology.

In the quantum regime was necessary to build one equivalent density
lagrangian in order to apply this, and does not possible to write this
solution in closed form. In this sense, we check this approach using
the original S\'aez-Ballester formalism, obtaining the exact solutions
in both regimes, classical and quantum for particular values in the
$\gamma$ parameter. This formalism will be used with anisotropic
cosmological models, which will be reported in other work.

\acknowledgments{This work was partially supported by CONACYT grants 47641, 56946 and  62253,
  DINPO 38.07 and PROMEP UGTO-CA-3. This work is part of the collaboration within the
  Instituto Avanzado de Cosmolog\'ia.}

\end{document}